\documentclass[aps,prd,twocolumn,floatfix]{revtex4}
\usepackage{amsmath,amssymb}
\usepackage{graphicx}

\begin{document}
\title{Results from the spin programme at COSY-ANKE}
\author{A.~Kacharava}
\affiliation{Institut f\"ur Kernphysik and J\"ulich Centre for Hadron Physics, Forschungszentrum J\"ulich, D-52425 J\"ulich, Germany}
\author{C.~Wilkin}
\affiliation{Physics and Astronomy Department, UCL, Gower Street, London WC1E 6BT, U.K.}
\begin{abstract}
Some of the important results from the COSY-J\"ulich spin
programme are summarised. These include the measurement of the
deuteron beam momentum through the excitation of a depolarising
resonance, which allowed the mass of the $\eta$-meson to be
determined to high precision. The charge exchange of polarised
deuterons on hydrogen gave rise to a detailed study of the spin
dependence of large angle neutron-proton elastic scattering
amplitudes. The measurements of the cross section and analysing
powers for pion production in both $pp$ and $pn$ collisions at
353~MeV could be described very successfully in terms of a
partial wave decomposition.
\end{abstract} \maketitle
%
%
\section{Introduction}
\label{sec1}

It was once claimed that in strong interaction physics ``Spin
is an inessential complication''~\cite{CHE1966}. This rather
negative attitude ignores certain practical applications and
surely dismisses the beauty inherent in many phenomena
involving particle polarisation. Recognising this, a programme
was outlined several years ago to exploit the possibilities of
carrying out experiments with polarised beams and targets at
ANKE~\cite{SPIN}. This facility is based around a magnetic
spectrometer sited at an internal target station of the COoler
SYnchrotron and storage ring COSY of the Forschungszentrum
J\"ulich. The basic features of the complex were described in
Ref.~\cite{BUS2011} and we shall here concentrate on a few of
the fruits of the spin programme.

COSY can accelerate and store polarised protons and vector and
tensor polarised deuterons up to momenta of 3.7~GeV/$c$. In
addition to unpolarised hydrogen and deuterium cluster-jet
targets, ANKE is also equipped with polarised $\vec{\rm H}$ and
$\vec{\rm D}$ gas target cells so that spin correlations can be
studied as well as beam and target analysing powers.

In the following three sections we first show how, in experiments with
polarised deuteron beams at a storage ring, the beam momentum can be
determined very precisely through the study of artificially induced
depolarising resonances. This led to a determination of the mass of the
$\eta$ meson that is as precise as any other in the literature. The
nucleon-nucleon programme has two distinct elements, the most developed being
the charge exchange of tensor polarised deuterons, which gives immediate
access to the tensor amplitudes in large angle neutron-proton scattering.
However, in addition, measurements are made in proton-proton elastic
scattering with polarised beam and target in angular regions where little
reliable data exist. Finally, a variety of spin-dependent data on pion
production in nucleon-nucleon collisions near threshold have been taken and
analysed. These permitted a full partial-wave analysis to be carried out in
the domain where two protons emerge almost bound.

%
%
\section{Beam momentum determination and the mass of the $\boldsymbol{\eta}$
meson}
\label{sec2}

A big challenge that one often faces in a precision experiment
at a storage ring is the determination of the beam momentum
with sufficient accuracy. Although the revolution frequency
$f_0$ can be measured with a relative precision of around
$10^{-5}$, there are much greater uncertainties in the exact
orbit of the particles in the ring. A way of overcoming this
problem was proposed many years ago~\cite{SER1976} and has
since been implemented at several electron colliders. Spin is
here very much the \emph{essential} element.

The spin of a polarised beam particle in a storage ring
precesses around the normal to the plane of the machine. A
horizontal \emph{rf} field from a solenoid can induce
depolarising resonances such that the beam depolarises when the
frequency of the externally applied field coincides with that
of the spin precession in the ring. The depolarising resonance
frequency $f_r$ depends upon the revolution frequency of the
machine and the kinematical factor $\gamma = E/mc^2$, where $E$
and $m$ are, respectively, the particle total energy and mass
of the particle. For a planar accelerator where there are no
horizontal fields,
\begin{equation}
\label{fr}
f_r/f_0=k + \gamma G\,,
\end{equation}
where $G$ is the particle's gyromagnetic anomaly and $k$ is an integer. The
combination of the measurements of the revolution and depolarising
frequencies allows the evaluation of $\gamma$ and hence the beam momentum
$p$.

The depolarising resonance technique was applied for the first
time at COSY with a vector polarised deuteron beam of momenta
around 3.1~GeV/$c$~\cite{GOS2010}. The deuterons were
accelerated with a \emph{rf} cavity and, once the required
momentum was reached, a barrier bucket cavity was used to
compensate for the energy losses incurred through the
beam-target interactions. The depolarising solenoid had an
integrated maximum longitudinal \emph{rf} magnetic field of
$\int B_{\text{rms}}\;d\ell =0.67$~T\,mm at a \emph{rf} voltage
of 5.7~kV rms. The value of $k=1$ in Eq.~(\ref{fr}) corresponds
to frequencies that were in the middle of the solenoid range of
0.5--1.5~MHz.

\begin{figure}[hbt]
\includegraphics[width=0.9\linewidth]{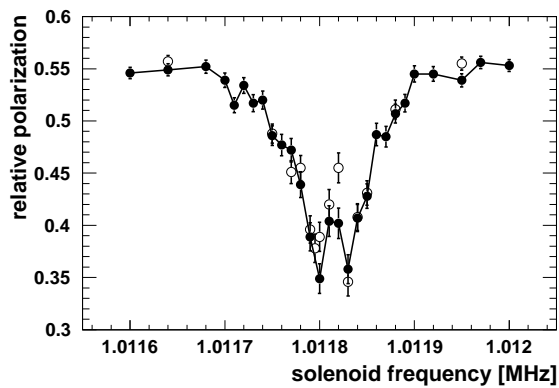}
\caption{\label{fig:fr_BegEnd}Spin-resonance measurements at a
single beam momentum (closed circles). The open symbols
represent results obtained for an extended cycle time, where
the perturbing solenoid was switched on after 178~s.}
\end{figure}

A vector polarised beam leads to an asymmetry in the scattering
from a carbon target, which could be measured with the EDDA
detector~\cite{ALT2005}. Since only the frequency of the
depolarising resonance needed to be determined, an absolute
calibration of this device at different deuteron momenta was
not required. Figure~\ref{fig:fr_BegEnd} displays an example of
this relative polarisation as a function of the solenoid
frequency for a fixed beam momentum. When the frequency of the
solenoid coincides with the spin-precession frequency, the beam
is maximally depolarised. The structures, especially the double
peak in the centre, are caused by the interaction of the
deuteron beam with the barrier bucket cavity. However, these
did not affect the mean position, which could be fixed with a
precision of $\approx 10^{-5}$. The full width at half maximum,
which was typically in the region of 80-100~Hz, is mainly a
reflection of the momentum spread within the beam. If this were
the only significant effect, it would correspond to $(\delta
p/p)_{\text{rms}} \approx 2 \times 10^{-4}$.

\begin{figure}[htb]
\includegraphics[width=0.9\linewidth]{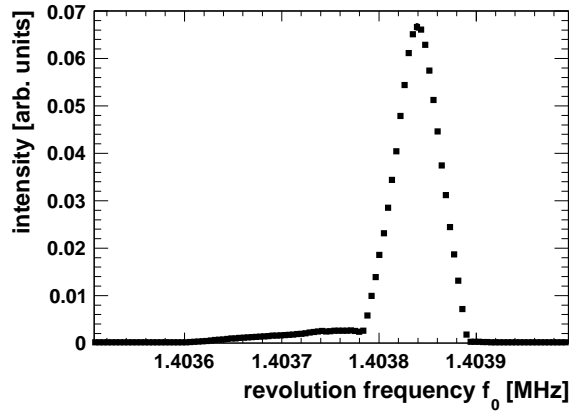}
\caption{\label{fig:f0_SC1FL1} Mean Schottky power spectrum at one beam
momentum. The statistical error bars lie within the data points. }
\end{figure}

The other frequency required in the evaluation of
Eq.~(\ref{fr}), i.e., that of the circulation in COSY, was
measured by using the Schottky noise of the deuteron beam. The
statistical distribution of the charged particles in the beam
leads to random current fluctuations that induce a voltage
signal at a beam pick-up in the ring. The Fourier transform of
this voltage-to-time signal by a spectrum analyser delivers the
frequency distribution around the harmonics of the revolution
frequency of the beam. As mentioned in Sec.~\ref{sec3}, this
phenomenon is also used at COSY to measure the luminosity in an
experiment~\cite{STE2008}. All the data acquired at a
particular beam momentum are presented in
Fig.~\ref{fig:f0_SC1FL1}. The small tail seen at low
frequencies corresponds to beam particles that escaped the
influence of the barrier bucket cavity but still circulated in
COSY. The statistical uncertainty in the weighted arithmetic
mean was in all cases below 0.2~Hz compared to the typical
1.4~MHz shown in the figure. This means that, under ideal
conditions, the left hand side of Eq.~(\ref{fr}) could be
measured with a precision of better than $10^{-5}$.

\begin{figure}[hbt]
\includegraphics[width=0.9\linewidth]{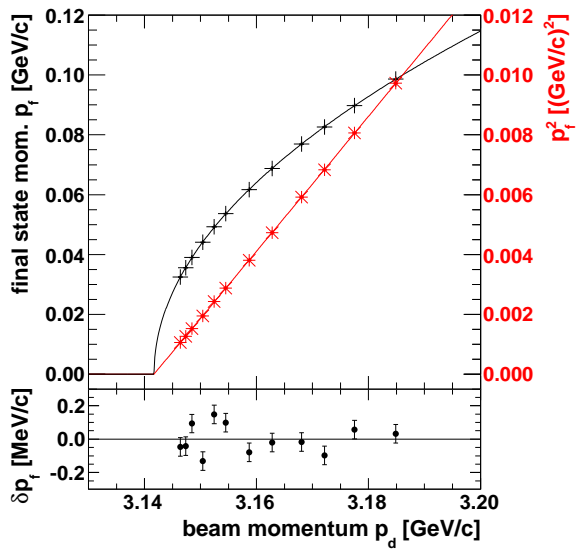}
\caption{\label{fig8} Values of the final-state CM momentum $p_f$ (black
crosses) and its square (red stars) plotted against the deuteron laboratory
momentum $p_d$. The lower panel shows the deviations of the experimental data
from the fitted curve in $p_f$.}
\end{figure}

The great efforts expended in determining precisely the
deuteron beam momentum were justified in order to measure the
mass of the $\eta$ meson from the missing-mass peak in the
$dp\to{}^{3}$He$\,X$ reaction~\cite{GOS2012}. For this purpose
the experiment was carried out at twelve closely spaced
deuteron momenta a little above the $\eta$ threshold and two
just below to provide the information required to subtract the
multipion background. By exploiting its full geometric
acceptance near threshold, it was possible to calibrate the
ANKE spectrometer very precisely and thus determine the final
$^3$He CM momentum $p_f$ for each of the twelve deuteron beam
momenta and the results are shown in Fig.~\ref{fig8}. Although
the method depends primarily upon the determination of the
kinematics rather than counting rates, its implementation is
helped enormously by the fact that the cross section jumps to
its plateau value already by the first point in
Fig.~\ref{fig8}~\cite{MER2007}.

The long lever arm facilitates a robust extrapolation to the
$\eta$ threshold, where the deuteron momentum was found to be
$p_d = 3141.686\pm0.021$~MeV/$c$. There is a one-to-one
relation between this and the mass of the meson, which is found
to be
\[m_{\eta}=(547.873\pm0.005_{\text{stat}}\pm0.026_{\text{syst}})~\text{MeV}/c^2.\]
It is in fact the determination of the threshold beam momentum
that provides the largest contribution to the 26~keV/$c^2$
systematic uncertainty. The result is compatible with all the
modern measurements reported by the Particle Data
Group~\cite{PDG2012} that studied the $\eta$ decay and the
error bars are as small as any of these. The result suggests
that earlier missing-mass determinations, which differed by
$\sim 0.5$~keV/$c^2$, lacked the necessary precision.

%
%
\section{The nucleon-nucleon programme}
\label{sec3}

A good understanding of the nucleon-nucleon ($NN$) interaction still remains
one of the principal goals of nuclear and hadronic physics. Apart from their
intrinsic importance for the study of nuclear forces, $NN$ elastic scattering
data are also necessary ingredients in the modeling of meson production and
other nuclear reactions at intermediate energies. It therefore goes without
saying that all facilities should try to fill in the remaining gaps in our
knowledge in the area.

The COSY-EDDA collaboration~\cite{ALT2005} produced a wealth of
data on proton-proton elastic scattering that completely
revolutionised the isospin $I=1$ $NN$ phase-shift analysis up
to about 2.1~GeV~\cite{SAID}. However, for proton energies
above about 1~GeV, very little is known about the $pp$ elastic
differential cross section or analysing power for
centre-of-mass angles $10^{\circ} <\theta_{\rm cm}<30^{\circ}$.
The cross section data that do exist seem to fall
systematically below the predictions of the SAID partial wave
analysis~\cite{SAID}. In this angular range the fast proton
emerging at small angles from a hydrogen target can be measured
well in the ANKE magnetic spectrometer, whereas the slow recoil
proton emerging at large angles can be measured independently
in one of the Silicon Tracking Telescopes. The luminosity that
is so crucial for the determination of the absolute cross
sections can be determined using the Schottky
method~\cite{STE2008} that was mentioned in the previous
section. Preliminary data are already available on the
differential cross sections at eight energies and approval has
been given to measure the proton analysing powers at the same
energies.

Much greater effort has been made in the study of the spin-dependent terms in
large angle neutron-proton scattering. It was pointed out many years ago that
the $dp\to \{pp\}_{\!s}n$ charge exchange at small angles was very sensitive
to the spin-spin terms in the $np\to pn$ amplitude provided the excitation
energy $E_{pp}$ in the final $pp$ system was kept low~\cite{BUG1987}. Under
such conditions the $\{pp\}_{\!s}$ is in a $^{1\!}S_0$ state and the charge
exchange necessarily involves a spin flip from the initial $np$ spin-triplet
of the deuteron. Furthermore, measurements of the deuteron tensor analysing
powers $A_{xx}$ and $A_{yy}$ allow one to distinguish between the
contributions from the three spin-spin $np$ amplitudes.

\begin{figure}[hbt]
\includegraphics[width=0.9\linewidth]{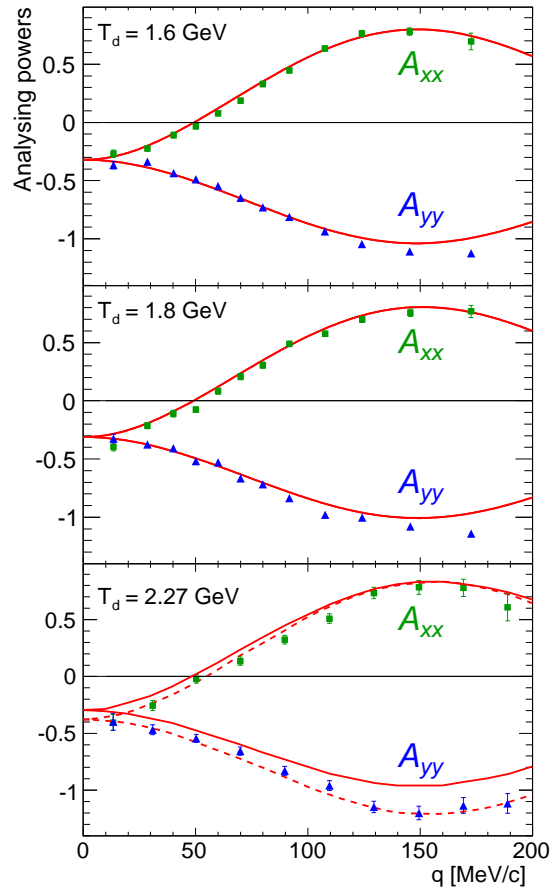}
\caption{\label{fig4} Cartesian deuteron analysing powers
for the $\vec{d}p \to \{pp\}_{\!s}n$ reaction for
$E_{pp}<3$~MeV at $T_d=1.2$, 1.6, and 1.8~GeV~\cite{MCH2012}
The impulse approximation predictions~\cite{CAR1991} have been
evaluated with the SAID amplitudes~\cite{SAID} (solid curves)
and also, at the highest energy, when the longitudinal
spin-spin amplitude is scaled by a factor of 0.75. }
\end{figure}

Measurements were carried out at Saclay~\cite{ELL1987,KOX1993}
but only in regions where the $NN$ amplitudes were reasonably
well known. These have been extended in fine steps in momentum
transfer $q$ to higher energy at ANKE~\cite{CHI2009,MCH2012}. A
cut of $E_{pp}<3$~MeV was typically imposed but any
contamination from triplet $P$-waves was taken into account in
the theoretical modelling~\cite{CAR1991}. The ANKE analysing
power results at 1.6, 1.8, and 2.27~GeV are compared in
Fig.~\ref{fig4} to these impulse approximation predictions
using up-to-date $np$ amplitudes~\cite{SAID} as input. The
satisfactory agreement at the two lower energies, and also in
the values of the differential cross sections, shows that the
theoretical description is adequate here.

Above about 1~GeV neutron-proton data become rather sparse. It
comes therefore as no surprise that, when the same approach is
employed on the higher energy data shown in Fig.~\ref{fig4},
the current SAID amplitudes~\cite{SAID} give a poor overall
description of the results. However, if the longitudinal
spin-spin amplitude is multiplied by a global factor of 0.75,
the agreement is much more satisfactory. This is evidence that
the charge exchange data can provide useful input to the $NN$
database.

\begin{figure}[hbt]
\includegraphics[width=0.9\linewidth]{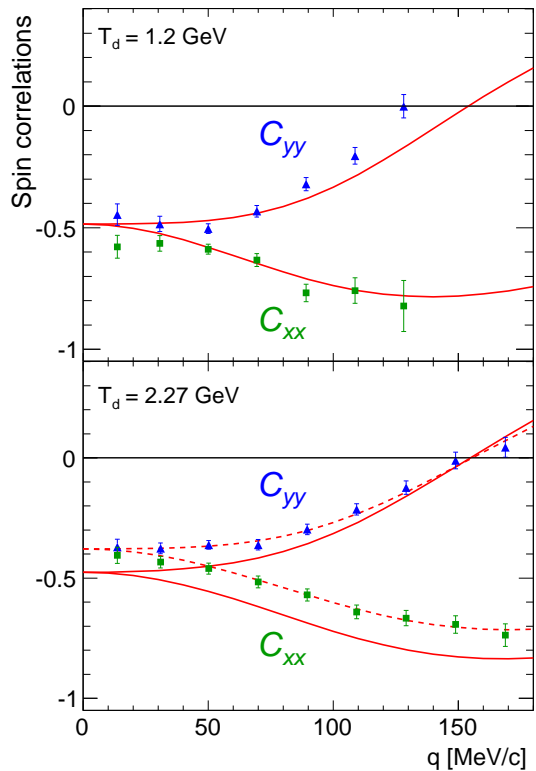}
\caption{\label{CxxCyy} Transverse spin correlation parameters
in the $\vec{d}\,\vec{p} \to \{pp\}_{\!s}n$ reaction at (a) 1.2
and (b) 2.27~GeV compared to the predictions of an impulse
approximation model (solid curves). Better agreement is found
at the higher energy if the longitudinal input is scaled by a
factor of 0.75 (dashed curves).}
\end{figure}

Confirmation of these conclusions is to be found in the
measurements of the deuteron-proton spin correlation parameters
measured with the polarised hydrogen gas cell. Results on this
are shown in Fig.~\ref{CxxCyy}. In impulse approximation, these
are sensitive to the interference between the longitudinal
spin-spin amplitude and the two transverse ones. Whereas there
is satisfactory agreement with the theoretical predictions at
1.2~GeV, the model is much more satisfactory at 2.27~GeV if the
longitudinal input is scaled by a factor of 0.75.

\begin{figure}[hbt]
\includegraphics[width=0.9\linewidth]{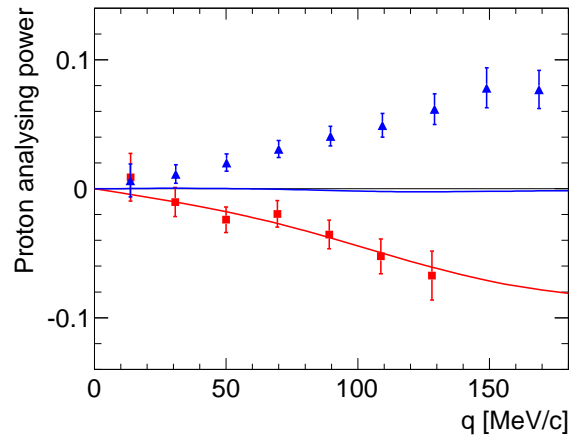}
\caption{\label{fig6} Proton analysing power in the
$d\vec{p}\to \{pp\}_{\!s}n$ reaction at 1.2~GeV (red squares)
and 2.27~GeV (blue triangles)~\cite{MCH2012} compared to
impulse approximation predictions. Note that, with the current
SAID input~\cite{SAID}, the latter almost vanish at the higher
energy.}
\end{figure}

In addition to measuring the spin correlations with the
polarised cell, data were also obtained on the proton analysing
power in the $d\vec{p}\to \{pp\}_{\!s}n$ reaction and the
results are shown in Fig.~\ref{fig6}. The message here is very
similar to that for the other observables. At 600~MeV per
nucleon the SAID input reproduces the experimental points very
well but it seems that at 1135~MeV the SAID description of the
spin-orbit amplitude has serious deficiencies.

\begin{figure}[hbt]
\includegraphics[width=0.9\linewidth]{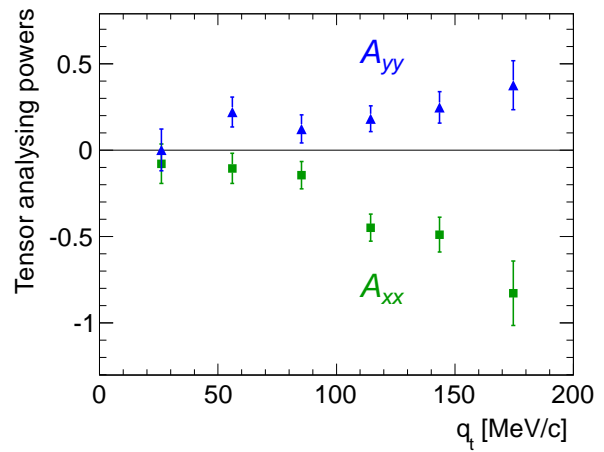}
\caption{\label{fig7}Tensor analysing powers for the
$\vec{d}p\to \{pp\}_{\!s}X$ reaction at 2.27~GeV as a function
of the transverse momentum transfer. The data are integrated
over the mass range $1.19 < M_X <
1.35$~GeV/$c^2$~\cite{MCH2011}. }
\end{figure}
\vspace{-2mm}

As well as studying the $\vec{d}p\to \{pp\}_{\!s}X$ data to
extract the neutron as a missing-mass peak, results were also
obtained where $m_X > m_N + m_{\pi}$. These events must be
associated with pion production, especially through the
$\Delta$ isobar. The first indications shown in Fig.~\ref{fig7}
are that the Cartesian analysing powers are largely opposite in
sign to those for $\vec{d}p \to \{pp\}_{\!s}n$~\cite{MCH2011}.
These data will therefore yield information on the amplitude
structure of the $NN\to N\Delta$ reaction.

%
%
\section{Pion production in nucleon-nucleon collisions}
\label{sec4}

One of the priorities at ANKE is to perform a complete set of measurements of
$NN\to\{pp\}_{\!s\,}\pi$ at low energy. Since, as in Sec.~\ref{sec3}, the
$\{pp\}_{\!s}$ proton-proton pair is overwhelmingly in the $^{1\!}S_0$ state,
only the polarisations of the initial nucleons have to be studied. As parts
of this programme, the differential cross section and analysing power of the
$\vec{p}p\to \{pp\}_{\!s\,}\pi^0$ reaction were measured at
353~MeV~\cite{TSI2012} and the same observables measured in quasi-free
$\pi^-$ production on the deuteron, $\vec{p}d\to p_{\rm sp}
\{pp\}_{\!s}\pi^-$~\cite{DYM2012}, where $p_{\rm sp}$ is a ``spectator''
proton. By making certain theoretical assumptions and retaining amplitudes up
to pion $d$-waves, the combined data sets are sufficient for a partial-wave
decomposition. This is of particular interest for Chiral Perturbation Theory,
where it is important to establish that the same short-ranged $NN\to NN\pi$
vertex that contributes to $p$-wave pion production is consistent with other
intermediate energy phenomena.

\begin{figure}[!ht]
\centering
\includegraphics[width=0.9\columnwidth]{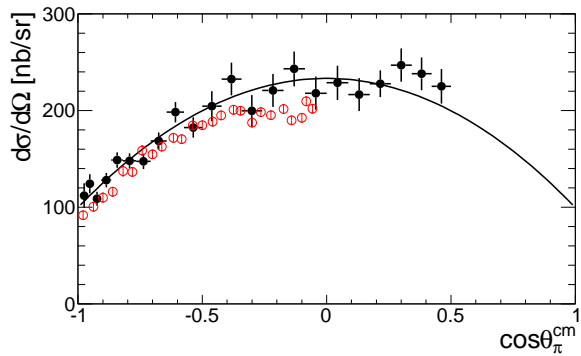}
\caption{\label{gig3} Differential cross section for the $pp\to
\{pp\}_{\!s\,}\pi^0$  reaction at 353~MeV. The ANKE measurements (solid
black) circles are compared with the CELSIUS data (open red) circles at
360~MeV~\cite{BIL2001}. The curve is the partial wave fit}
\end{figure}

\begin{figure}[!ht]
\centering
\includegraphics[width=0.9\columnwidth]{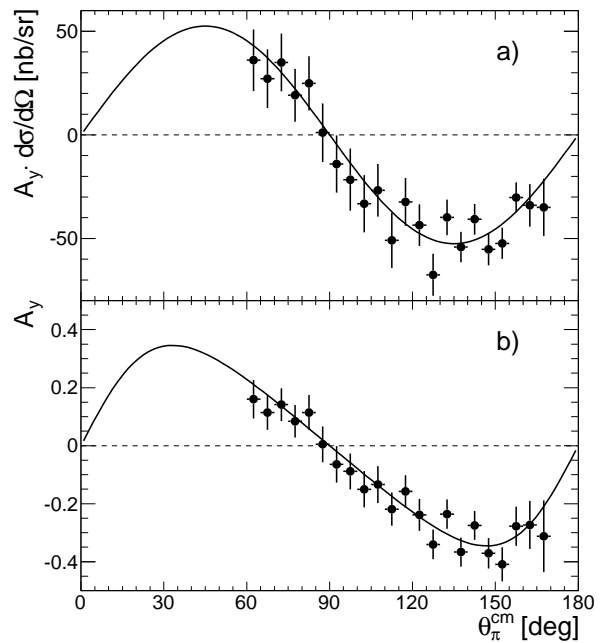}
\caption{\label{gig4} (a) Product of the measured analysing power and
differential cross section for the $\vec{p}p\to \{pp\}_{\!s\,}\pi^0$
reaction. (b) Measured values of $A_y$; the overall systematic uncertainty is
$\approx 5$\%. The curves are partial wave fits.}
\end{figure}

For $\pi^0$ production, both protons were measured in the ANKE Forward
Detector. After selecting the $^{1\!}S_0$ final state, the kinematics of the
$pp\to\{pp\}_{\!s}X$ process could be reconstructed on an event-by-event
basis to obtain the $\pi^0$ rate from the missing-mass $M_{\!X}$ spectrum. By
using a beam with a $\pm68$\% polarisation, the cross section and analysing
power could be measured simultaneously and the results are shown in
Figs.~\ref{gig3} and \ref{gig4}.

The cross section data agree quite well over most of the angular range with
those taken at CELSIUS~\cite{BIL2001} and the strong anisotropy is evidence
for significant $d$-wave pion production. In the absence of pion $d$- (or
higher) waves the analysing power would vanish and, as seen in
Fig.~\ref{gig4}, this is far from being the case.

\begin{figure}[htb]
\vspace*{1mm}
\centering
\includegraphics*[width=0.9\columnwidth]{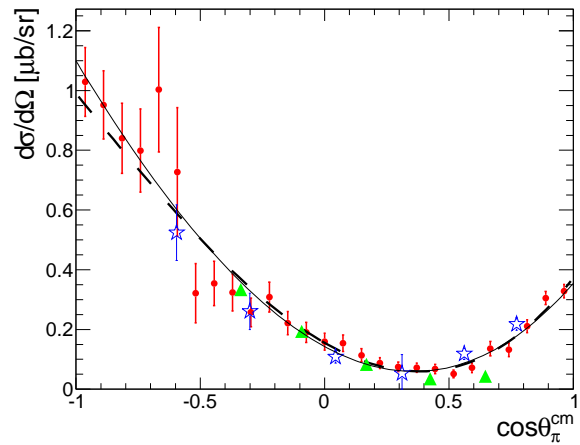}
\caption{Differential cross section for $pn\to \{pp\}_{\!s\,}\pi^-$ at
$T\approx 353$~MeV. ANKE data with statistical errors are shown by red
circles; the systematic error is $\approx 6$\%. The statistical errors of the
TRIUMF data~\cite{DUN1998} (green triangles) are smaller than the symbols and
the normalisation uncertainty is 10\%. The blue stars are arbitrarily scaled
cross sections extracted from pion absorption data~\cite{HAH1996}. The solid
curve is a partial wave fit.} \label{fig:DSG}
\end{figure}

\begin{figure}[!ht]
\centering
\includegraphics[width=0.9\columnwidth]{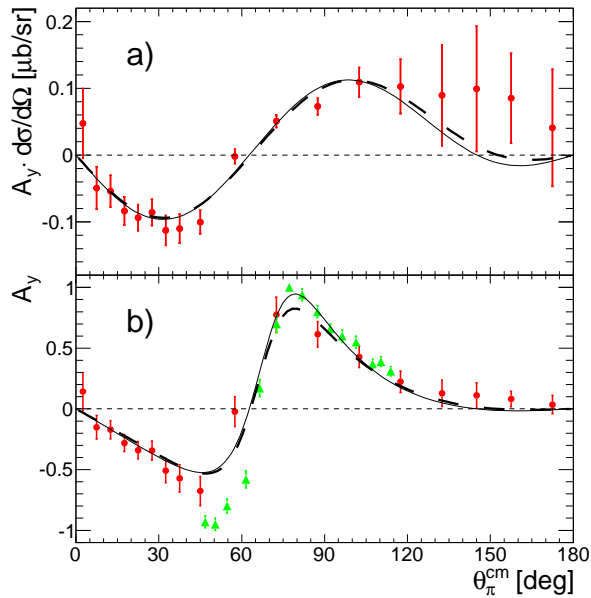}
\caption{\label{fig:DSGAy} (a) Product of the measured
analysing power and differential cross section for the
$\vec{p}n\to \{pp\}_{\!s\,}\pi^-$ reaction at 353~MeV; the
statistical error do not include the 11\% systematic
uncertainty. (b) Values of $A_y$ measured at ANKE (circles) and
TRIUMF~\cite{HAH1999} (triangles). The solid curves are partial
wave fits to the ANKE data. }
\end{figure}

In the $\vec{p}d\to p\{pp\}_{\!s}\pi^-$ experiment, three
particles had to be detected in the final state to identify the
reaction. In addition to the two protons in the $^{1\!}S_0$
state, either the $\pi^-$ or the third (slow) proton must be
measured, the latter in one of the silicon tracking telescopes
placed in the target chamber. Together the two detection modes
led to a full angular coverage. In either case the slow proton
was restricted kinematically to be a spectator so that the
cross section and analysing power of the quasi-free
$\vec{p}n\to \{pp\}_{\!s\,}\pi^-$ reaction could be extracted
in the $353\pm20$~MeV interval, the results being shown in
Figs.~\ref{fig:DSG} and \ref{fig:DSGAy}.

The differential cross section agrees with the earlier TRIUMF
measurement~\cite{DUN1998}, except for their two most forward points. The
disagreement persists with the analysing power data measured in the forward
hemisphere~\cite{HAH1999} shown in Fig.~\ref{fig:DSGAy}. On the other hand,
the agreement with the shape of the cross section deduced from the
$\pi^-{}^{3}\textrm{He}\to pnn$ reaction~\cite{HAH1996} is even better.

Even if one considers only $s$, $p$, and $d$-wave pion production, the cross
section and analysing power data are insufficient to perform a full amplitude
analysis without further assumptions. These were to neglect the coupling
between the initial $^{3\!}P_2$ and $^{3\!}F_2$ waves and to use the Watson
theorem to determine the phases of the production amplitudes from these and
also the $^{3\!}P_0$  wave. There are then seven real parameters available to
describe essentially ten features in Figs.~\ref{gig3}~--~\ref{fig:DSGAy}. The
success achieved here suggests that the phase assumptions are basically
correct. The analysis shows that $d$-wave production is confined almost
purely to the $^{3\!}P_2$ channel but by far the largest term is associated
with $p$-wave production from the initial $^{3\!}D_1$ state.

%
%

\section{The future}
\label{sec5}

Although the partial wave description of the pion production
data is both plausible and impressive, one needs to measure
other types of observables in order to test its validity. By
using a polarised deuterium gas cell in conjunction with a
polarised proton beam, it was possible to study the transverse
spin-spin correlation in the
$\vec{p}\vec{n}\to\{pp\}_{\!s\,}\pi^-$ reaction. The
preliminary results are consistent with the predictions of the
amplitude analysis discussed in Sec.~\ref{sec4}. Further checks
could be made through measurements of the
longitudinal-transverse spin correlation but these will require
the delivery, installation, and commissioning of a Siberian
snake to rotate the proton spin. This should take place early
in 2013. The snake will also allow us to study the
spin-correlation parameter $A_{00kn}$ in small angle $pp$
elastic scattering.

Though the charge exchange programme with a polarised deuteron
beam has been very successful, this only allows measurements to
be carried out up to 1.15~GeV per nucleon. To go higher at COSY
we must work in inverse kinematics and use the polarised
deuterium target in conjunction with a proton beam. The charge
exchange can then be studied purely through the measurement of
two slow protons in the silicon tracking telescopes without
using the ANKE magnetic spectrometer at all. However, this
opens even more fascinating possibilities, such as the study of
$\Delta$ isobar production in $\vec{p}\,\vec{d}\to
\{pp\}_{\!s}\Delta^0$, where the spin alignment of the $\Delta$
isobar can be determined through the measurement of one of the
products of the $\Delta^0\to p\pi^-$ decay. On the other hand,
with its array of detectors, ANKE can investigate
simultaneously a wide range of nuclear reactions, which makes
the spin programme at the facility so exciting.

%

\end{document}